\documentclass[11pt]{cernrep}
\usepackage{graphicx}
\usepackage{ulem}
\usepackage{epsfig}

\newcommand{\simr}{\hspace{.3em}\raisebox{.4ex}{$>$}\hspace{-.87em}
 \raisebox{-.7ex}{$\sim$}\hspace{.3em}}

\begin{document}

\title{Sign problem and MEM\thanks{~Poster presented by
Y. Shinno. E-mail:{\tt shinno@th.phys.saga-u.ac.jp}}
\thanks{~SAGA-HE-225, YGHP-06-37}}

\author{Masahiro Imachi$^1$, Yasuhiko Shinno$^2$ and Hiroshi Yoneyama$^2$}

\institute{$^1$Department of Physics, Yamagata University, Yamagata
990-8560, Japan \\
$^2$ Department of Physics, Saga University, Saga 840-8502, Japan}

%

\maketitle

\begin{abstract}
The sign problem is notorious in Monte
 Carlo simulations of lattice QCD with the finite density, lattice field
 theory (LFT) with a $\theta$ term and quantum spin models. In this
 report, to deal with the sign problem, we apply the maximum entropy
 method (MEM) to LFT with the $\theta$ term and investigate to what extent the
 MEM is applicable to this issue. Based on this study, we also make a
 brief comment about lattice QCD with the finite density in terms of the
 MEM. 
\end{abstract}

\section{Introduction}
It is an important subject to reveal the phase structure of QCD in
$\mu$-$T$ space, where $T$ and $\mu$ are temperature and quark chemical
potential, respectively. This gives hints not only to understand the
physics of 
the early universe and the neutron star, but also to analyze what
happens in heavy ion collisions. The lattice simulation is
one of the most reliable methods to comprehensively study the phase
structure. However, Monte Carlo (MC) simulation based on the importance
sampling method cannot directly
apply to Lattice QCD at the finite density, because the
fermion determinant with $\mu$ makes the Boltzmann weight
complex. This is the notorious sign problem. Although  various
techniques to circumvent this problem have been
proposed,\cite{rf:Schmidt} the sign 
problem has not been solved yet. In this report, the maximum entropy
method (MEM)\cite{rf:Bryan,rf:JG} is introduced from a different
viewpoint. By applying 
the MEM to lattice field theory (LFT) with a $\theta$ term, where it
also suffers from the sign problem, we investigate to
what extent the MEM is applicable to this issue. Based on this study, we
make a brief comment about lattice QCD at the finite density in terms of
the MEM. 

\section{Sign Problem in LFT with the $\theta$ Term}
The partition function ${\cal Z}(\theta)$ in LFT with the $\theta$ term
can be calculated by 
Fourier-transforming the topological charge distribution $P(Q)$:
\begin{equation}
 {\cal Z}(\theta)=\frac{\int[d{\bar z}dz]
 e^{-S({\bar z},z)+i\theta{\hat Q}({\bar z},z)}}
 {\int[d{\bar z}dz]e^{-S({\bar z},z)}}
  \equiv \sum_Q e^{i\theta Q}P(Q), \label{eqn:partitionfunction}
\end{equation}
 where $S(\bar{z},z)$ and ${\hat Q}(\bar{z},z)$ are the action and the
 topological charge as functions of lattice fields $\bar{z}$ and
 $z$, respectively. Note that $P(Q)$ is calculated with a real positive
 Boltzmann  weight. We call 
 this the Fourier transform method (FTM). Although
 this method works well for small volumes, it breaks down for large
 volumes. This is because the error in $P(Q)$ disturbs the behavior of
 the free energy density $f(\theta)\equiv-\frac{1}{V}\log{\cal
 Z}(\theta)$ ($V$ is a volume).  
\begin{figure}[t]
\begin{center}
\includegraphics[width=80mm]{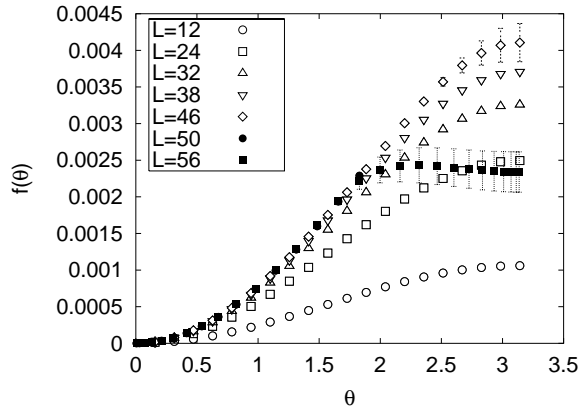}
\end{center}
\vskip -5mm
\caption{
 Free energy density $f(\theta)$ obtained from the MC data of the CP$^3$
 model. The coupling $\beta$ is fixed to 3.0 and lattice
 sizes $L$ are changed from 12 to 56. The number of measurements reaches
 several millions for each case.}
\label{fig:CP3FP-Fourier}
\end{figure}
Figure \ref{fig:CP3FP-Fourier} displays $f(\theta)$ obtained from MC
data of the CP$^3$ model with the fixed point action. The coupling
$\beta$ is fixed to 3.0 and various lattice sizes $L$ are employed. The
number of measurements is several millions for each case. Although $f(\theta)$
for $L\leq 38$ behaves smoothly in the whole $\theta$ region,
$f(\theta)$ for $L=50$ and 56 cannot be properly calculated for
$\theta\simr 2.0$. In the $L=56$ case, $f(\theta)$ becomes flat for
$\theta\simr 2.0$. This is called flattening.  In the $L=50$ case,
$f(\theta)$ cannot be obtained for $\theta\simr 2.0$ due to negative
values of ${\cal Z}(\theta)$.  We also call it flattening, because the
error in $P(Q)$ causes this behavior in the same way as the $L=56$
case. Flattening is originated from the sign problem. This is understood
in the following way.\cite{rf:PS,rf:IKY}  The MC data of 
$P(Q)$ consists of 
the true value of $P(Q)$, ${\tilde P}(Q)$, and its error, $\Delta P(Q)$.
When the error in $P(Q)$ at $Q=0$ dominates, $f(\theta)$ is approximated
by $f(\theta)\simeq -\frac{1}{V}\log[e^{-V{\tilde
f}(\theta)}+\Delta P(0)]$. Here,
${\tilde f}(\theta)$ denotes the true value of $f(\theta)$. Since
$f(\theta)$ is an increasing function of $\theta$, $e^{V{\tilde
f}(\theta)}\simeq |\Delta P(0)|$ could occur at $\theta=\theta_{\rm f}$ and
$f(\theta)\simeq -\frac{1}{V}\log\delta P(0)$ for $\theta\simr
\theta_{\rm f}$. To overcome this problem
requires the number of measurements proportional to $e^V$.

\section{MEM}
The MEM is one of the parameter inference based on Bayes' theorem and
derives a unique solution by utilizing data and our knowledge about the
parameters.\cite{rf:Bryan,rf:JG,rf:AHN} In our MEM
analysis,\cite{rf:ISY,rf:ISY2} the inverse Fourier transform 
\begin{equation}
 P(Q)=\int^{\pi}_{-\pi}d\theta\frac{e^{-i\theta Q}}{2\pi}{\cal Z}(\theta).
 \label{eqn:invFourier}
\end{equation}
is used. The MEM involves to maximize the posterior
probability ${\rm prob}({\cal Z}(\theta)|P(Q),I)$.  Here, \\
${\rm prob}({\cal Z}(\theta)|P(Q),I)$ is the probability that ${\cal
Z}(\theta)$ is realized when the MC data of $\{P(Q)\}$ and information
$I$ are given. Information $I$ represents our state of knowledge about
${\cal Z}(\theta)$ and ${\cal Z}(\theta)>0$ is imposed. The probability
is given by 
\begin{equation}
 {\rm prob}({\cal Z}(\theta)|P(Q),I)\propto \exp\left[-\frac{1}{2}\chi^2
		+\alpha S\right]\equiv e^{W[{\cal Z}]}, \label{eqn:posterior}
\end{equation}
where $\chi^2$, $\alpha$ and $S$ denote a standard $\chi^2$-function, a
real positive parameter and an entropy, respectively. Conventionally, the
Shannon-Jaynes entropy
\begin{equation}
 S=\int_{-\pi}^{\pi} d\theta\left[{\cal Z}(\theta)-m(\theta)
	-{\cal Z}(\theta)\log\frac{{\cal Z}(\theta)}{m(\theta)}\right]
\label{eqn:S-Jentropy}
\end{equation}
is employed. A function $m(\theta)$ is called default model and is chosen
so as to be consistent with $I$. The most probable image ${\hat{\cal
Z}}(\theta)$ is obtained
according to the following procedures. (1) To obtain the most probable
image for a given $\alpha$, ${\cal Z}^{(\alpha)}(\theta)$, by maximizing
$W[{\cal Z}]$. (2) To obtain the $\alpha$-independent most probable
image ${\hat{\cal Z}}(\theta)$ by averaging ${\cal
Z}^{(\alpha)}(\theta)$ over $\alpha$; ${\hat{\cal Z}}(\theta)=\int
d\alpha~P(\alpha){\cal Z}^{(\alpha)}(\theta)$. The probability
$P(\alpha)$ represents the posterior probability of $\alpha$. (3) To
estimate the error in ${\hat{\cal Z}}(\theta)$ as the uncertainty of
${\hat{\cal Z}}(\theta)$. The probability $P(\alpha)$ is given by  
$P(\alpha)\propto g(\alpha)e^{W(\alpha)+\Lambda(\alpha)}$. 
Here, $W(\alpha)\equiv W[{\cal Z}^{(\alpha)}]$, and $\Lambda(\alpha)$
represents contributions of fluctuations of ${\cal Z}(\theta)$ around
${\cal Z}^{(\alpha)}(\theta)$. The function $g(\alpha)$ is the prior
probability of $\alpha$. Conventionally, two types of $g(\alpha)$ are
used: $g_{\rm 
Lap}(\alpha)={\rm const}$ (Laplace's rule) and $g_{\rm
Jef}(\alpha)=1/\alpha$ (Jeffrey's rule). Information about
$\alpha$ before obtaining data does not play the conclusive role in the
derivation of ${\hat{\cal Z}}(\theta)$. 
In the present study, the
$g(\alpha)$-dependence of ${\hat{\cal Z}}(\theta)$ is estimated by the
following quantity:
\begin{equation}
 \Delta(\theta)\equiv 
  \frac{|{\hat{\cal Z}}_{\rm Lap}(\theta)-{\hat{\cal Z}}_{\rm Jef}(\theta)|}
  {{\hat{\cal Z}}_{\rm Lap}(\theta)}, \label{eqn:Delta}
\end{equation}
where ${\hat{\cal Z}}_{\rm Lap}(\theta)$ and ${\hat{\cal Z}}_{\rm
Jef}(\theta)$ are the most probable images for Laplace's and
Jeffrey's rules, respectively. 

\section{Numerical Results}
We apply the MEM to the MC data with
flattening as well as without flattening. The latter is the data for
$L=38$ (data A) and the 
former is those for $L=50$ (data B). Here, two types of $m(\theta)$ are
used: (i) Gaussian function $m_{\rm G}(\theta)=\exp[-\gamma\frac{\log
10}{\pi}\theta^2]$, where a parameter $\gamma$ is changed over a
wide range, and (ii) $m(\theta)={\hat{\cal Z}}(\theta)$ for smaller
volumes. In case (ii), to analyze the data for $L=L_0$, ${\hat{\cal
Z}}(\theta)$ obtained by the MEM for smaller volumes are utilized as
$m(\theta)$. 
For $L_0=50$, ${\hat{\cal Z}}(\theta)$ for $L=24$, 32 and 38
are used as $m(\theta)$, which are denoted as
$m_{L/L_0}(\theta)=m_{L/50}(\theta)$. In this report, all results of the
MEM with Laplace's rule are shown  except for $\Delta(\theta)$. In the
analysis, the Newton method 
with quadruple precision is used. 

\subsection{Non Flattening Case}
\begin{figure}[h]
\begin{center}
\includegraphics[width=80mm]{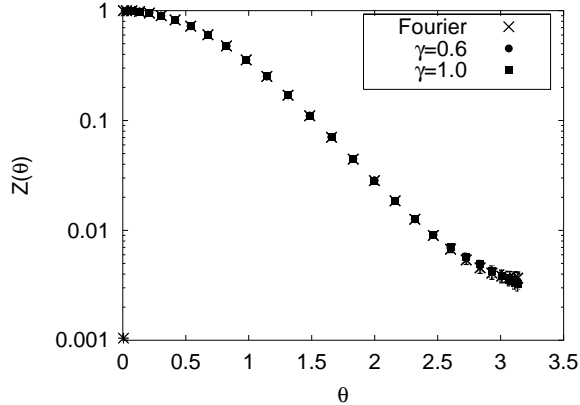}
\end{center}
\vskip -8mm
\caption{The most probable images in the non flattening case
 ($L=38$). The result of the FTM is also plotted ($\times$).}
\label{fig:finalZ_L38}
\end{figure}

Figure \ref{fig:finalZ_L38} displays ${\hat{\cal Z}}(\theta)$ for
data A.  The Gaussian defaults $m_{\rm G}(\theta)$ with $\gamma=0.6$ and
1.0 are 
used. The partition function ${\cal Z}_{\rm Four}(\theta)$ obtained by
the FTM is also plotted. Both the results of the MEM have no
$m(\theta)$-dependence and are in 
agreement with the result of the FTM. The
error of ${\hat{\cal Z}}(\theta)$,  $\delta{\hat{\cal Z}}(\theta)$, are
calculated according 
to the procedure (3). These errors are too small to be visible in
Fig. \ref{fig:finalZ_L38}.

\subsection{Flattening Case}
\begin{figure}
\includegraphics[width=75mm]{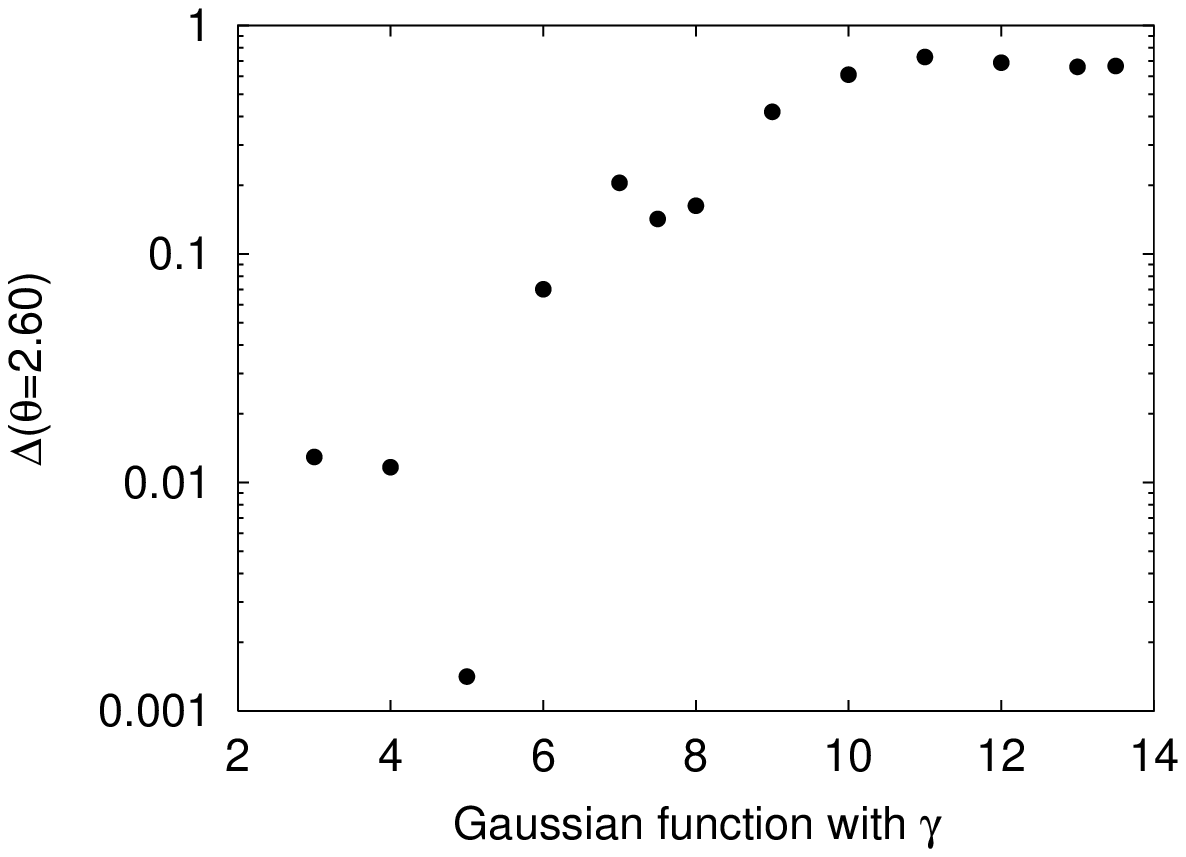}
\includegraphics[width=75mm]{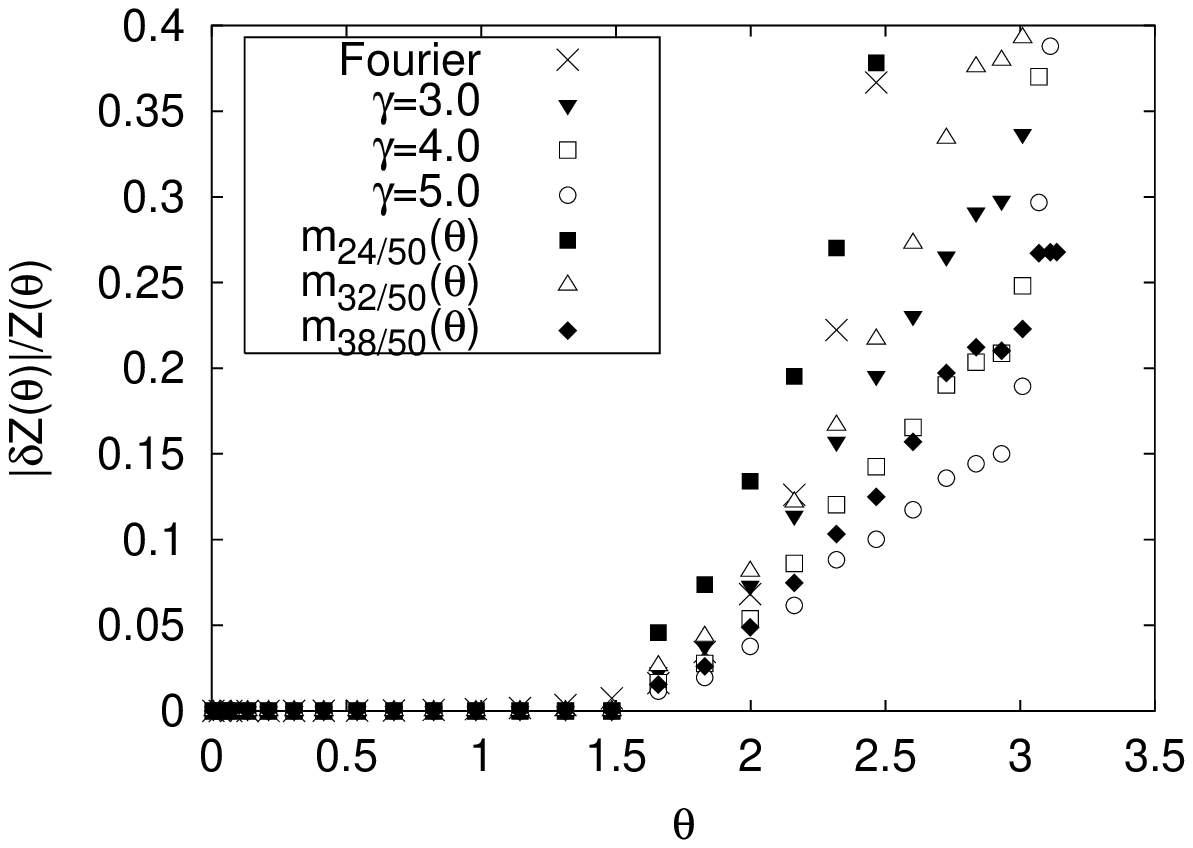}
\vskip -5mm
\caption{Values of $\Delta(\theta)$ at $\theta=2.60$ (left panel). The
 Gaussian defaults are used. Values of 
 $|\delta{\hat{\cal Z}}(\theta)|/{\hat{\cal Z}}(\theta)$ for the selected
 6 default models (right panel). The value of 
$|\delta{\cal Z}_{\rm Four}(\theta)|/{\cal Z}_{\rm Four}(\theta)$ is also
 plotted in the right panel.} 
\label{fig:constraints}
\end{figure}

\begin{figure}[h]
\includegraphics[width=75mm]{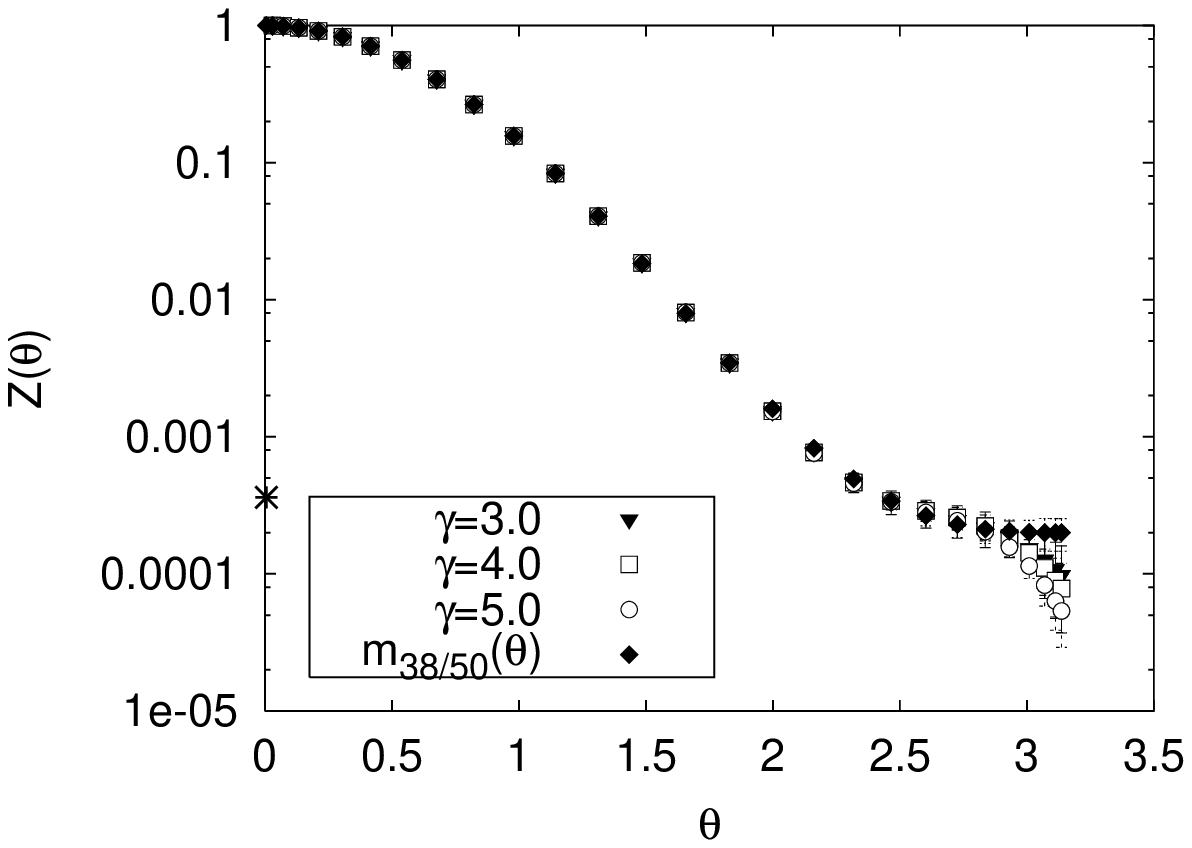}
\includegraphics[width=75mm]{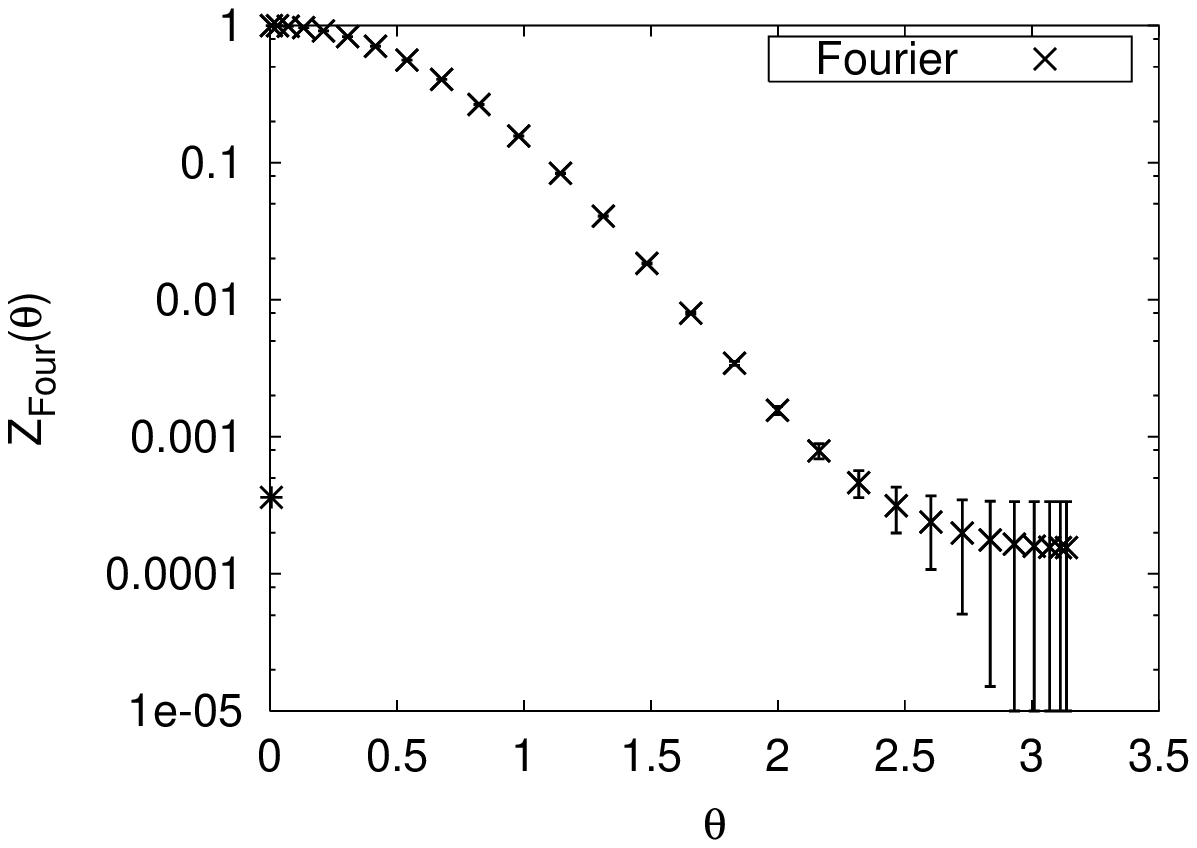}
\vskip -5mm
\caption{The most probable images for the selected $m(\theta)$ (left 
 panel). As a comparison, ${\cal Z}_{\rm Four}(\theta)$ obtained by the
 FTM is also displayed in the right panel. The total number of
 measurements is 30.0M/set for both the cases.} 
\label{fig:finalZ_L50}
\end{figure}

Let us turn to data B. Unlike data A, much care is needed in the
analysis.\cite{rf:ISY} In order to properly evaluate ${\hat{\cal
Z}}(\theta)$ as the final image, we investigate (i) the statistical
fluctuation of
${\hat{\cal Z}}(\theta)$, (ii) $g(\alpha)$-dependence of ${\hat{\cal
Z}}(\theta)$ and (iii) the relative error of ${\hat{\cal
Z}}(\theta)$. In (i), it is found that the statistical fluctuation
of ${\hat{\cal Z}}(\theta)$ becomes smaller with increasing the number
of measurements and that ${\hat{\cal Z}}(\theta)$ with 20.0M/set is
obtained with sufficiently small fluctuations 
except for near $\theta=\pi$. In (ii), we systematically
investigate the $g(\alpha)$-dependence of ${\hat{\cal Z}}(\theta)$ by
calculating $\Delta(\theta)$. The left panel of
Fig. \ref{fig:constraints} displays 
$\Delta(\theta)$ at $\theta=2.60$, as a representative. Here, the
Gaussian defaults are used, where 
$\gamma$ is changed from 3.0 to 13.5. The value of $\Delta(\theta)$ is
the smallest for $\gamma=5.0$ and becomes larger as the value of
$\gamma$ deviates from 5.0. Similar results are obtained in the whole
$\theta$ region. This seems to indicate that $m_{\rm G}(\theta)$ with
$\gamma=5.0$ is the most suitable as $m(\theta)$ among the defaults
which we have chosen. Keeping in mind that
$\Delta(\theta)$ includes an uncertainty originated from
$\delta{\hat{\cal Z}}(\theta)$, we impose a
constraint that the final images should satisfy 
$\Delta(\theta)<0.2$. Here, this value is chosen as a typical one of the
uncertainty in $\Delta(\theta)$ coming from $\delta{\hat{\cal
Z}}(\theta)$. Six images 
satisfy this constraint among those which we have obtained, and 
do not depend on $m(\theta)$ up to $\theta=3.0$. In (iii),
we investigate how the MEM is applicable to our issue by calculating the
relative error $|\delta{\hat{\cal Z}}(\theta)|/{\hat{\cal
Z}}(\theta)$. Upon a constraint $|\delta{\hat{\cal
Z}}(\theta)|/{\hat{\cal Z}}(\theta)<0.3$, the four most probable images
${\hat{\cal Z}}(\theta)$ satisfy the constraint up to $\theta=3.0$. This
constraint is chosen from the fact that the error propagation of $P(Q)$
starts to strongly affect the behavior of ${\cal Z}_{\rm Four}(\theta)$
at $|\delta{\cal Z}_{\rm Four}(\theta)|/{\cal Z}_{\rm
Four}(\theta)\simeq 0.3$ in the FTM (see the right panel of
Fig. \ref{fig:finalZ_L50}). Here, this value realizes at smaller value
of $\theta$, $\theta=2.4$. These results are displayed in
the right panel of Fig. \ref{fig:constraints}. \\
In this analysis, we
find  that the four most probable
images are obtained with reasonably small errors in a wide range of
$\theta$, which is displayed in the left panel of
Fig. \ref{fig:finalZ_L50}. As a comparison, ${\cal Z}_{\rm
Four}(\theta)$ is also shown in the right panel.

\section{Summary and Discussions}
In this report, to deal with the sign problem in LFT with the $\theta$
term, we apply the MEM to the MC data of the CP$^3$ model. In non
flattening case, all results of the MEM agree with the one of the FTM
within the errors. In the flattening case, obtained images depend on
$m(\theta)$. By investigating whether they are adequate images, we have
found that the MEM allows us to 
calculate ${\cal Z}(\theta)$ with small errors for large $\theta$
region. For the details, see Ref. \cite{rf:ISY3}\\
Finally, let us make a brief comment about lattice QCD with the finite
density in terms of the MEM. In lattice QCD with the finite chemical
potential, MC simulation cannot be directly performed due to
the complex phase of the fermion determinant. 
There are various techniques to avoid the sign problem and we
concentrate on the canonical ensemble
approach.\cite{rf:RW,rf:HT,rf:AKW,rf:AFHL} By the fugacity expansion,
${\cal Z}(V,T,\mu)$ is written as
\begin{equation}
 {\cal Z}(V,T,\mu)=\sum_n Z(V,T,n)(e^{\mu/T})^n, \label{eqn:fugacity}
\end{equation}
where $n$ is the total quark number. Taking $\mu=i\phi T$, where $\phi$
is a real, ${\cal 
Z}(V,T,\mu=i\phi T)$ is free from the sign problem and 
${\cal Z}(V,T,\mu=i\phi T)$, in principle, can be calculated with MC
simulation. In 
this case, ${\cal Z}(V,T,\mu=i\mu T)=\sum_n Z(V,T,n)e^{in\phi}$.
Comparing it 
with Eq. (\ref{eqn:invFourier}), we see the following correspondence:
\begin{equation}
 \{P(Q)\leftrightarrow{\cal Z}(V,T,\mu=i\phi T),~
  e^{-i\theta Q}/2\pi\leftrightarrow e^{i\phi n},~
 {\cal Z}(\theta)\leftrightarrow Z(V,T,n)\}. 
 \label{eqn:correspondence}
\end{equation}
It may be worthwhile to study the theory in terms of the MEM.

\section*{Acknowledgments}
The authors thank R. Burkhalter for providing his FORTRAN code for the
CP$^{N-1}$ model with the fixed point action.
One of the authors (Y. S.) is also grateful to G. Akemann for fruitful
information.  This work is supported
in part by Grants-in-Aid for Scientific Research (C)(2) of the Japan
Society for the Promotion of Science (No. 15540249) and of the Ministry
of Education Science, Sports and Culture (No's 13135213 and
13135217). Numerical calculations have been performed on the computer
at Computer and Network Center, Saga University.



\begin{thebibliography}{99}
\bibitem{rf:Schmidt} C. Schmidt, {\tt hep-lat/0408047} and references
	in this paper.
\bibitem{rf:Bryan} R. K. Bryan, Eur. Biophys. J. {\bf 18} (1990), 165.
\bibitem{rf:JG} M. Jarrell and J. E. Gubarnatis, Phys. Rep. {\bf 269}
	(1996), 133.
\bibitem{rf:PS} J. C. Plefka and S. Samuel, Phys. Rev. {\bf D56} (1997),
	44.
\bibitem{rf:IKY} M. Imachi, S. Kanou and H. Yoneyama,
	Prog. Theor. Phys. {\bf 102} (1999), 653.
\bibitem{rf:AHN} M. Asakawa, T. Hatsuda and Y. Nakahara,
	Prog. Part. Nucl. Phys. {\bf 46} (2001), 459.
\bibitem{rf:ISY} M. Imachi, Y. Shinno and H. Yoneyama,
	Prog. Theor. Phys. {\bf 111} (2004), 387.
\bibitem{rf:ISY2} M. Imachi, Y. Shinno and H. Yoneyama,  {\tt
	hep-lat/0506032} 
\bibitem{rf:ISY3} M. Imachi, Y. Shinno and H. Yoneyama, in progress.
\bibitem{rf:RW} A. Roberge and N. Weise, Nucl. Phys. B {\bf 275} (1986),
	734.
\bibitem{rf:HT} A. Hasenfratz and D. Toussaint, Nucl. Phys. B {\bf 371}
	(1992), 539.
\bibitem{rf:AKW} M. Alford, A. Kapustin and F. Wilzek, Phys. Rev. D {\bf
	59} (1999), 05402.
\bibitem{rf:AFHL} A. Alexandru, N. Faber, I. Horv\'ath
	and K.-F. Liu, {\tt hep-lat/0507020}.

\end{thebibliography}
\end{document}